\documentclass[10pt,twocolumn,prl,aps,amssymb,amsmath,showpacs,tightenlines]{revtex4}

\usepackage{dsfont}
\usepackage[dvips]{graphicx}
\usepackage{color}
\usepackage{subfigure}
\usepackage{listings}

\newcommand{\half}{\mbox{$\textstyle \frac{1}{2}$}}
\newcommand{\re}{\mbox{$\rm e$}}
\newcommand{\ri}{\mbox{$\rm i$}}
\newcommand{\rd}{\mbox{$\rm d$}}

\begin{document}

\title{Quantum splines}

\author{Dorje~C.~Brody${}^1$, Darryl~D.~Holm${}^2$, David~M.~Meier${}^2$}

\affiliation{${}^1$Mathematical Sciences, Brunel University, Uxbridge UB8 3PH, UK \\ 
${}^2$Department of Mathematics, Imperial College London, London SW7 2AZ, UK}


\begin{abstract}
A quantum spline is a smooth curve parameterised by time in the space of unitary transformations, 
whose associated orbit on the space of pure states traverses a designated set of quantum states at 
designated times, such that the trace norm of the time rate of change of the associated Hamiltonian 
is minimised. The solution to the quantum spline problem is obtained, and is applied in an example 
that illustrates quantum control of coherent states. An efficient numerical scheme for computing 
quantum splines is discussed and implemented in the examples. 
\end{abstract}


\maketitle

Controlling the evolution of the unitary transformations that generate quantum dynamics is vital in 
quantum information processing. There is a substantial literature devoted to the investigation of 
the many aspects of quantum control \cite{QC}. The objective of quantum control is the unitary 
transformation of one quantum state, pure or mixed, into another one, subject to certain criteria. 
For example, one may wish to transform a given quantum state $|\psi\rangle$ into another state 
$|\phi\rangle$ unitarily in the shortest possible time, with finite energy resource 
\cite{DCB,Hosoya,BH}. When only the initial and final states are involved, many time-independent 
Hamiltonians are available that achieve the unitary evolution $|\psi\rangle\to|\phi\rangle$, 
and we simply need to find one that is optimal. However, transforming a given quantum state 
$|\psi\rangle$ along a path that traverses through a sequence of designated quantum states 
$|\psi\rangle\to|\phi_1\rangle\to |\phi_2\rangle\to\cdots\to|\phi_n\rangle$ cannot be achieved by a 
time-independent Hamiltonian. To realise this chain of transformations in the shortest possible time, 
one chooses the optimal Hamiltonian $H_j$ for each interval $|\phi_j\rangle\to|\phi_{j+1}
\rangle$ \cite{Hosoya,BH}, and switches the Hamiltonian from $H_j$ to $H_{j+1}$ when the state has 
reached $|\phi_{j+1}\rangle$. However, instantaneous switching of the Hamiltonian is in general not 
experimentally feasible. 

In the present paper, we consider the following quantum control problem: Let a set of quantum 
states $|\phi_1\rangle$, $|\phi_2\rangle$, $\cdots$, $|\phi_m\rangle$ and a set of times $t_1$, 
$t_2$, $\cdots$, $t_m$ be given. Starting from an initial state $|\psi_0\rangle$ at time $t_0=0$, 
find a time-dependent Hamiltonian $H(t)$ such that the evolution path $|\psi_t\rangle$ passes 
arbitrarily close to $|\phi_{j}\rangle$ at time $t = t_{j}$ for all $j=1,\ldots,m$, and such that the 
change in the Hamiltonian, in a sense defined below, is minimised. The solution to this problem 
will generate a continuous curve in the space of quantum states that interpolates through the 
designated states, just as a spline curve interpolates through a given set of data points. We 
thus refer to this solution as a \textit{quantum spline}. 

\begin{figure}[t!]
\begin{center}
\includegraphics[scale=0.10]{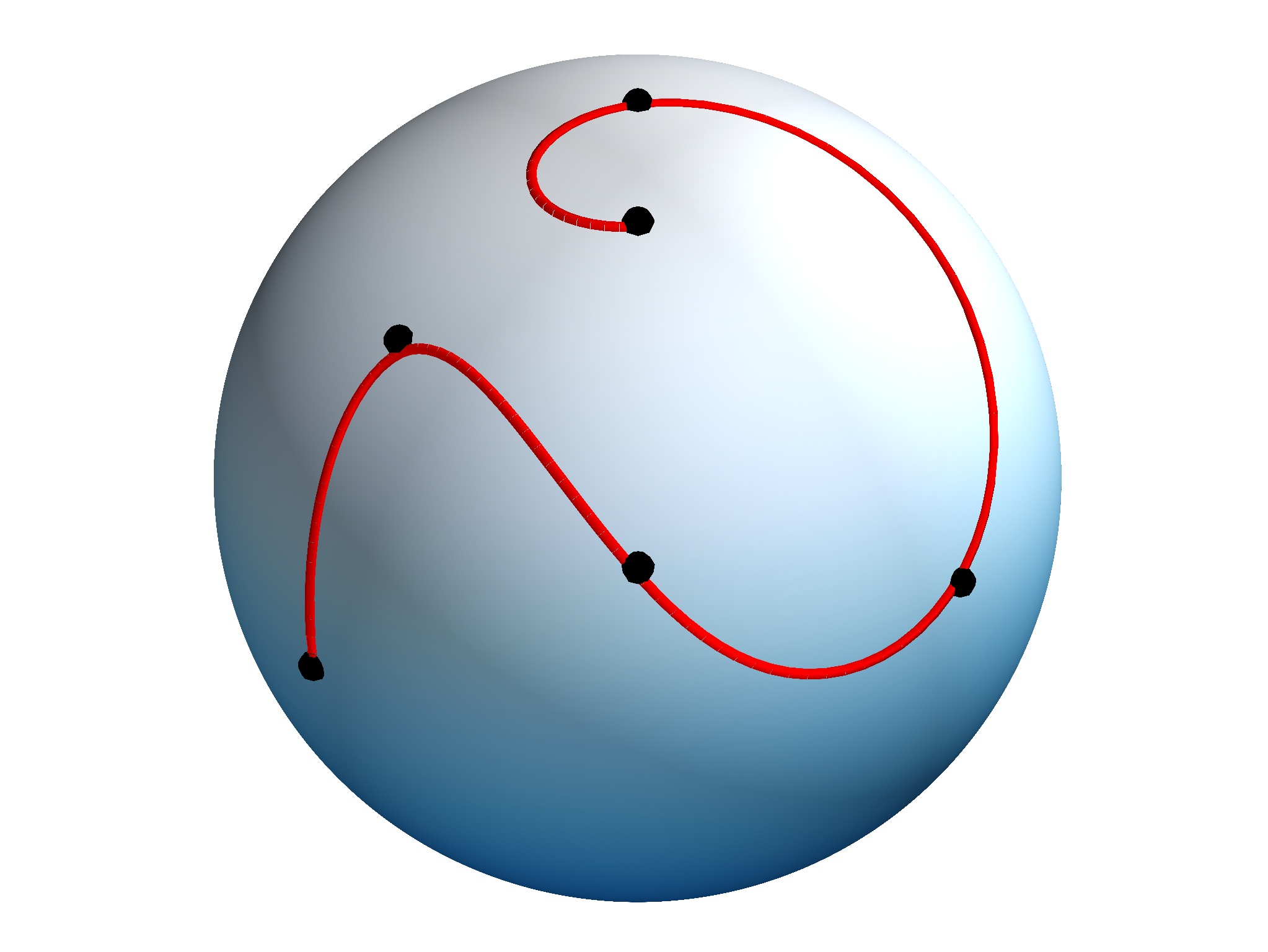}
\footnotesize
\caption{(colour online) 
\textit{A quantum spline for a two-level system}. The lower-left initial state and the 
targets are represented by black dots. The variational formulation of the problem requires 
to minimise a functional that measures both the cost related to the change of the 
Hamiltonian, and the amount of mismatch between the trajectory and the target points.
\label{fig:1}}
\end{center}
\end{figure}

There is a difference between 
a classical spline curve and a quantum spline. In the classical context the solution curve passes 
through a given set of points, whereas in the quantum context, a curve on the space of pure 
states in itself has no operational meaning. Thus, instead of finding a curve in the space of pure states 
where the designated  states lie, we must find a time-dependent curve in the space of Hamiltonians 
that in turn will generate the curve in the unitary transformation group needed to produce an optimal 
trajectory. In other words, we shall seek a curve in the associated Lie algebra, which of course 
is equivalent to the space of Hamiltonians, up to multiplication by $\ri=\sqrt{-1}$. 

Our approach involves variational calculus in the Lie algebra of 
skew-Hermitian matrices, with constraints that take values in the unitary group. In addition, since 
our optimality condition for quantum splines involves the time-derivative of $\ri H(t)$, we shall make 
use of the techniques developed recently for the higher-order calculus of variations on Lie groups 
and their algebras~\cite{GHM,BHM}. By extending these results we are able to: (a) derive the 
Euler-Lagrange equations (\ref{eq:5}) and (\ref{eq:9}) below that solve quantum spline problems; and 
(b) devise an efficient discretisation scheme to numerically implement the solution. An example of 
such a solution for a two-level quantum system is sketched in Fig.~\ref{fig:1}. As an application, 
we illustrate how the results transform a quantum state along a path that lies entirely on the 
coherent-state subspace. 

The optimal curve $H(t)$ that solves the quantum spline problem is the minimiser of a `cost 
functional' (action) consisting of two terms: The first term measures the overall change in the 
Hamiltonian during the evolution. For this purpose we shall consider the trace norm, i.e. for a 
pair of trace-free skew-Hermitian matrices $A$ and $B$ we define their inner product by 
\begin{eqnarray}
\langle A, B \rangle = -2\, {\rm tr}(AB),
\label{eq:1}
\end{eqnarray}
where the factor $-2$ is purely conventional. Thus, if $H$ is a time-dependent Hamiltonian 
and ${\dot H}$ its time derivative, the instantaneous penalty arising from changing the 
Hamiltonian is given by $\frac{1}{2}\langle\ri{\dot H},\ri{\dot H}\rangle={\rm tr}({\dot H}^2)$. The 
second term penalises the `mismatch' between the state $|\psi_{t_j}\rangle$ at time $t_j$ and 
the target state $|\phi_j\rangle$. For this purpose we shall use the standard geodesic 
distance:
\begin{eqnarray}
D(\psi,\phi) = 2 \arccos \sqrt{ \frac{\langle\psi|\phi\rangle\langle\phi|\psi\rangle}
{\langle\psi| \psi\rangle \langle\phi|\phi\rangle}}
\label{eq:2}
\end{eqnarray}
for a pair of states $|\psi\rangle$ and $|\phi\rangle$. Writing $U(t)$ for the parametric family of 
unitary operators generated by $H(t)$ so that $|\psi_{t_j}\rangle = U(t_j)|\psi_0\rangle$, the 
mismatch penalty is chosen to be $\frac{1}{2}D^2(U(t_j)\psi_0,\phi_j)/\sigma^2$,where the 
tolerance $\sigma>0$ is a tunable parameter so that the penalty is high when $\sigma$ is 
small, and the factor of a half is purely conventional. 

The action, of course, must be minimised subject to the constraint that the dynamical 
evolution of the state is unitary. That is, $U$ must satisfy the Schr\"odinger equation 
${\dot U}=-\ri H U$, in units $\hbar=1$. Therefore, given an initial state $|\psi_0\rangle$ at 
time $t_0=0$, a set of target states $|\phi_{1}\rangle$, $\cdots$, $|\phi_m\rangle$ at times 
$t_1, \cdots, t_m$, and an initial Hamiltonian $H(0) = H_0$, we wish to find the minimiser of 
\begin{eqnarray}
{\mathcal J} &=& \int_{t_0}^{t_m} \!\Big( \half \langle \ri {\dot H}, \ri {\dot H}\rangle 
+ \langle M, \dot{U}U^{-1} + \ri H \rangle \Big) \rd t \nonumber \\ && + 
\frac{1}{2\sigma^2} \sum_{j=1}^m D^2(U(t_j)\psi_0,\phi_j)\,, 
\label{eq:3} 
\end{eqnarray}
where the minimisation is over curves $U(t)\in SU(n+1)$ and $\ri H(t),\, M(t) 
\in\mathfrak{su}(n+1)$. Additionally, we require smoothness of these curves on 
open intervals $(t_j, t_{j+1})$ for $j=0,\ldots,m-1$; $U(0) = {\mathds 1}$; and the 
continuity of $U(t)$ and $H(t)$ is assumed everywhere. The curve $M(t)$ acts 
as a Lagrange multiplier enforcing the kinematic constraint. 

Before we proceed to vary the action ${\mathcal J}$ let us comment on the choice 
of the initial Hamiltonian $H_0$. We let $H_0$ be such that the trajectory 
$\re^{-{\rm i}H_0t}|\psi_0\rangle$ corresponds to the geodesic curve on the space of pure 
states joining $|\psi_0\rangle$ and $|\phi_1\rangle$; the construction of such a Hamiltonian 
can be found in \cite{BH}. Intuitively, since the first target time $t_1$ is fixed, this 
choice generates the most direct traverse $|\psi_0\rangle\to|\phi_1\rangle$, hence 
requiring least change in the Hamiltonian at initial times $t\ll t_1$. 

The Euler-Lagrange equations governing stationary points of (\ref{eq:3}) are obtained by 
taking the variation of ${\mathcal J}$ and requiring $\delta \mathcal{J}=0$. Writing 
$A = (\delta U) U^{-1}$ we have 
\begin{eqnarray}
\delta{\mathcal J} &=& \int_{t_0}^{t_m} \Big( \langle \ri\dot{H}, \ri\delta \dot{H}\rangle + 
\langle M, \dot{A} - [\dot{U}U^{-1}, A] + \ri \delta H\rangle \nonumber \\ && 
+ \langle\delta M, \dot{U}U^{-1} + \ri H\rangle\Big) \rd t +  \frac{1}{2\sigma^2} 
\sum_{j=1}^m \delta D^2(\psi_{t_j}, \phi_j) \nonumber \\ &=& \int_{t_0}^{t_m} \Big( 
\langle M - \ri\ddot{H}, \ri\delta H\rangle + \langle -\dot{M} + [\dot{U}U^{-1}, M], A\rangle 
\nonumber \\ && +  \langle\delta M, \dot{U}U^{-1} + \ri H\rangle\Big) \rd t  + 
\frac{1}{2\sigma^2} \sum_{j=1}^{m} \delta D^2(\psi_{t_j}, \phi_j) \nonumber \\ && 
+ \sum_{j = 1}^{m-1} \left[\langle \Delta M(t_j), A(t_j)\rangle + \langle 
\ri\Delta {\dot H}(t_j), \ri\delta H(t_j)\rangle \right]\nonumber \\ && 
+ \langle M(t_m), A(t_m)\rangle + \langle \ri\dot{H}(t_m), \ri\delta H(t_m)\rangle,  
\label{eq:4}
\end{eqnarray}
where in the second step we have integrated by parts, and used the notations 
$\Delta M(t_j)=M(t_j^-) - M(t_j^+)$ and $\Delta {\dot H}(t_j)={\dot H}(t_j^-) - 
{\dot H}(t_j^+)$, with $M(t_i^+)=\lim_{t \downarrow t_i}M(t)$ and 
$M(t_i^-)=\lim_{t\uparrow t_i}M(t)$; and similarly for ${\dot H}(t_j^\pm)$. It 
follows from (\ref{eq:4})  that on the open intervals $(t_j, t_{j+1})$, 
$j = 0, \ldots, m-1$, the following equations hold:
\begin{eqnarray}
\ri{\ddot H} - M = 0, \quad \!\!\!\!{\dot M} + [M, \dot{U}U^{-1}] = 0, \quad \!\!\!\!
{\dot U}U^{-1} + \ri H = 0. 
\label{eq:5}
\end{eqnarray}
Additionally, at the nodes $t=t_j$, we require matching conditions. To work them out, 
let us calculate the variation $\delta D^2=2D\delta D$ appearing in (\ref{eq:4}). From 
the definition (\ref{eq:2}) and the relation 
\begin{eqnarray}
\frac{\langle\psi|\re^{-\varepsilon A}|\phi\rangle\langle\phi|\re^{\varepsilon A}|\psi\rangle}
{\langle\psi| \psi\rangle \langle\phi|\phi\rangle} &\approx& 
\frac{\langle\psi|(1-\varepsilon A)|\phi\rangle\langle\phi|(1+\varepsilon A)|\psi\rangle}{\langle\psi|
\psi\rangle \langle\phi|\phi\rangle} \nonumber \\ 
&& \hspace{-3.0cm} = \frac{\langle\psi|\phi\rangle\langle\phi|\psi\rangle}{\langle\psi|
\psi\rangle \langle\phi|\phi\rangle}  + \frac{2\Re[\langle\psi|\phi\rangle\langle\phi|A
|\psi\rangle]}{\langle\phi|\phi\rangle \langle\psi|\psi\rangle}\varepsilon  + \mathcal{O}(\varepsilon^2),
\label{eq:6}
\end{eqnarray}
which holds for any $A=-A^\dagger$, we find, bearing in mind that if 
$D=2\arccos(\sqrt{x})$ then $\rd D/\rd x = -2/\sin(D)$, 
\begin{eqnarray}
\delta D &=&  \left.\frac{\rd }{\rd \varepsilon} D( \re^{\varepsilon A}\psi,\phi) \right|_{\varepsilon = 0} = 
\frac{-4 \Re[ \langle \psi|\phi\rangle \langle \phi|A|\psi \rangle]}{\sin(D) \langle \phi|\phi \rangle \langle 
\psi|\psi\rangle}.  
\label{eq:7}
\end{eqnarray}
From (\ref{eq:7}), and writing $D_j=D(\psi_{t_j}, \phi_j)$, we deduce that $\delta D_j^2 = 
2D_j \langle F_j, A(t_j)\rangle$, where 
\begin{eqnarray}
F_j = \frac{\langle\psi_{t_j}|\phi_j\rangle |\psi_{t_j}\rangle \langle\phi_j| -
\langle\phi_j|\psi_{t_j}\rangle |\phi_j\rangle \langle\psi_{t_j}|}
{\sin (D_j)\langle\phi_j|\phi_j\rangle \langle\psi_{t_j}|\psi_{t_j}\rangle}.
\label{eq:8}
\end{eqnarray} 
The relevant matching conditions at the nodes are therefore given by: 
\begin{eqnarray} 
{\dot H}(t_j^+)-{\dot H}(t_j^-)=0, \quad 
 M(t_j^+)-M(t_j^-) = D_j F_j/\sigma^2 , 
 \label{eq:9}
\end{eqnarray}
whereas we require ${\dot H}(t_m) = 0$ and $M(t_m) + D_m F_m/\sigma^2=0$ at the terminal 
point. Quantum spline problems are therefore solved by finding a solution to equations 
(\ref{eq:5}) and (\ref{eq:9}) that satisfies, in addition, the terminal conditions at $t_m$. 
On open time intervals $(t_i, t_{i+1})$ equation (\ref{eq:5}) yields 
\begin{eqnarray}
\dddot{H} + \ri [H, \ddot{H}] = 0.
\label{eq:10}
\end{eqnarray}
This is the right-reduced equation for the so-called Riemannian cubics on $SU(n+1)$ with 
respect to the bi-invariant Riemannian metric induced by the inner product \eqref{eq:1}. 
That is, $U(t)$ is a Riemannian cubic on the open time intervals $(t_i, t_{i+1})$. Here, 
by a \textit{Riemannian cubic} we mean a solution to a certain fourth-order equation for a curve 
on a Riemannian manifold (see \cite{NHP1989} for further details). The node conditions 
\eqref{eq:9} imply that $U(t)$ is a \textit{Riemannian cubic spline}, a twice continuously 
differentiable curve that is composed of a series of cubics.

\begin{figure}[t]
\centering
\subfigure[$\sigma = 0.04$]{
\includegraphics[scale=0.1]{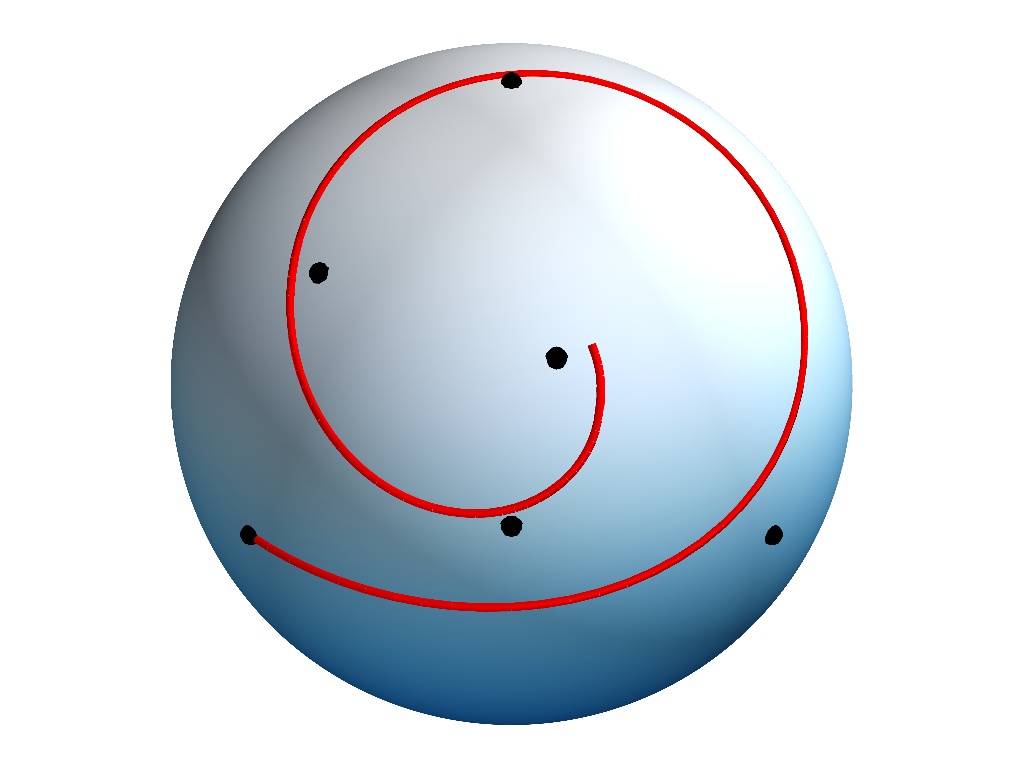}
\label{QSpline:subfig1}}
\qquad
\subfigure[$\sigma = 0.01$]{
\includegraphics[scale=0.1]{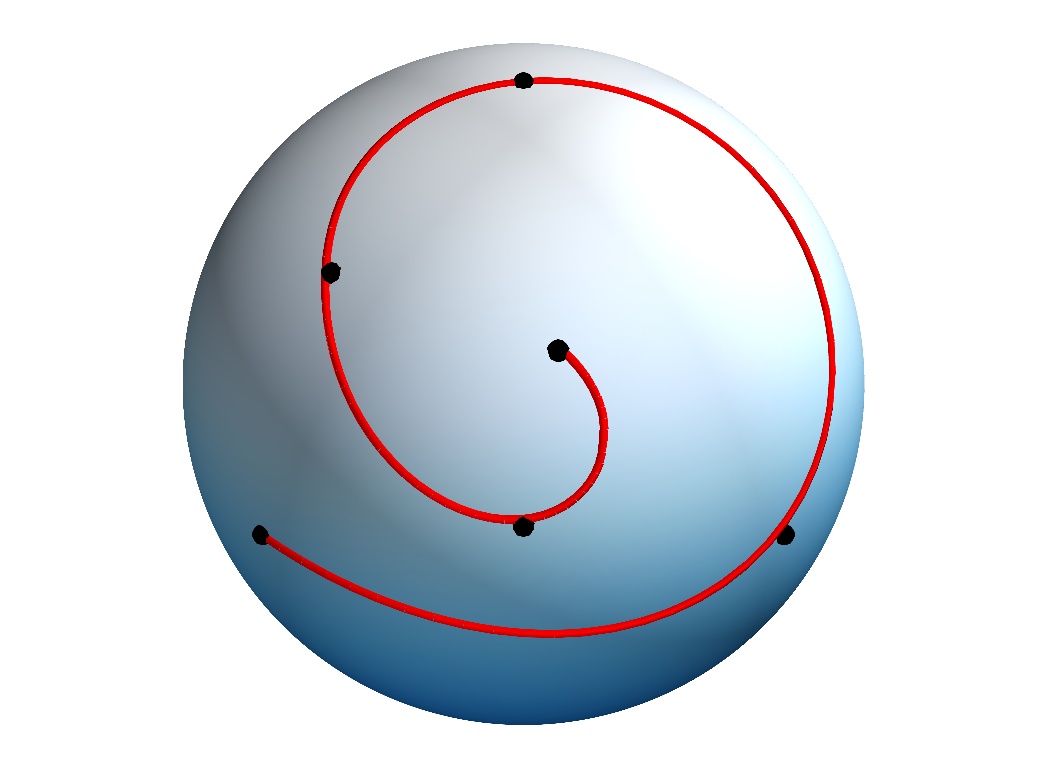}
\label{QSpline:subfig2}}
\caption{\footnotesize (Colour online) 
\textit{Orbits on the state space generated by the solution to the quantum spline problem}. 
The black dots indicate the initial (lower left) and the target points. 
The optimal trajectories are shown for two different values of the tolerance parameter: 
$\sigma=0.04$ and $\sigma=0.01$. Lower values of the tolerance parameter translate, through 
the cost functional ${\mathcal J}$, into a stronger penalty on the mismatch. 
\label{fig:2}
}
\end{figure}

We remark on the important structure of the Lagrange multiplier $M(t)$ implied by the 
equations of motion that makes it sufficient to 
consider a subspace of $\mathfrak{su}(n+1)$ when searching for the 
optimal initial value $M(0)$. Let us denote by ${\mathcal P}_\psi$ the totality of trace-free 
skew-Hermitian 
generators of unitary motions that leave the state $|\psi\rangle$ invariant, and 
${\mathcal P}_\psi^\perp$ its complement with respect to the inner product \eqref{eq:1}. Then, 
we have the following \textit{Lemma}: $M(t)\in {\mathcal P}_{\psi_t}^\perp$ (this holds because 
for all $j$, $D_jF_j\in\mathcal{P}_{\psi_{t_j}}^\perp$, and from (\ref{eq:5}), $M(t)_{t\in(t_j, t_{j+1})} = 
{\rm Ad}_{U(t)U(t_{j+1})^{-1}} M(t_{j+1}^-)$). 
This result is significant, because the search for the optimal $M(0)$ can be restricted to 
the $2n$-dimensional subspace $\mathcal{P}_{\psi_0}^\perp$ of the $n(n+2)$-dimensional 
Lie algebra $\mathfrak{su}(n+1)$.

\begin{figure}[t]
\centering
\subfigure[Evolution of the rotation axis ${\boldsymbol n}(t)$ for $\sigma = 0.04$]{
\includegraphics[scale=0.05]{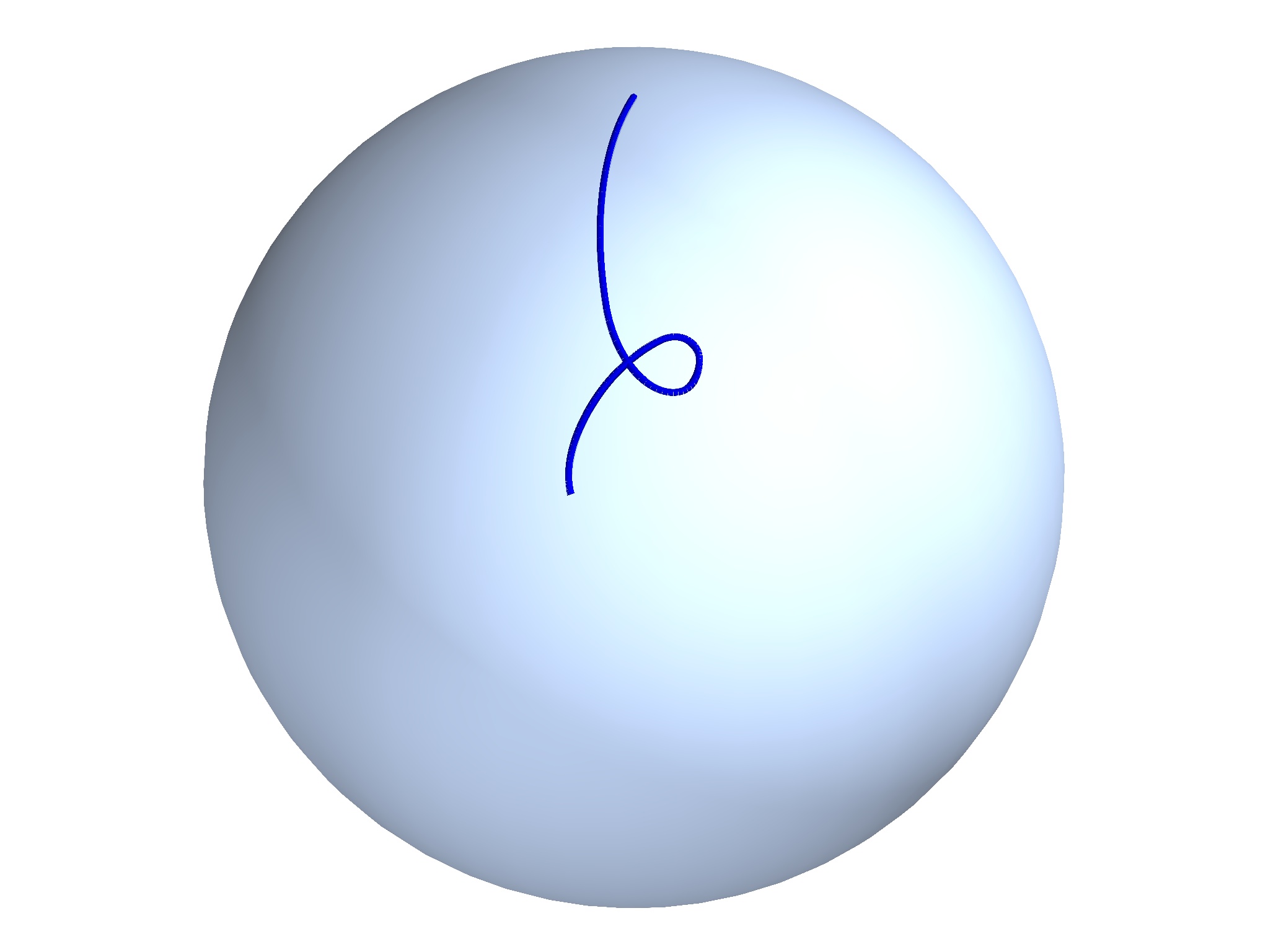}
\label{HCurves:subfig1}}
\qquad
\subfigure[Evolution of the rotation axis ${\boldsymbol n}(t)$ for  $\sigma = 0.01$]{
\includegraphics[scale=0.05]{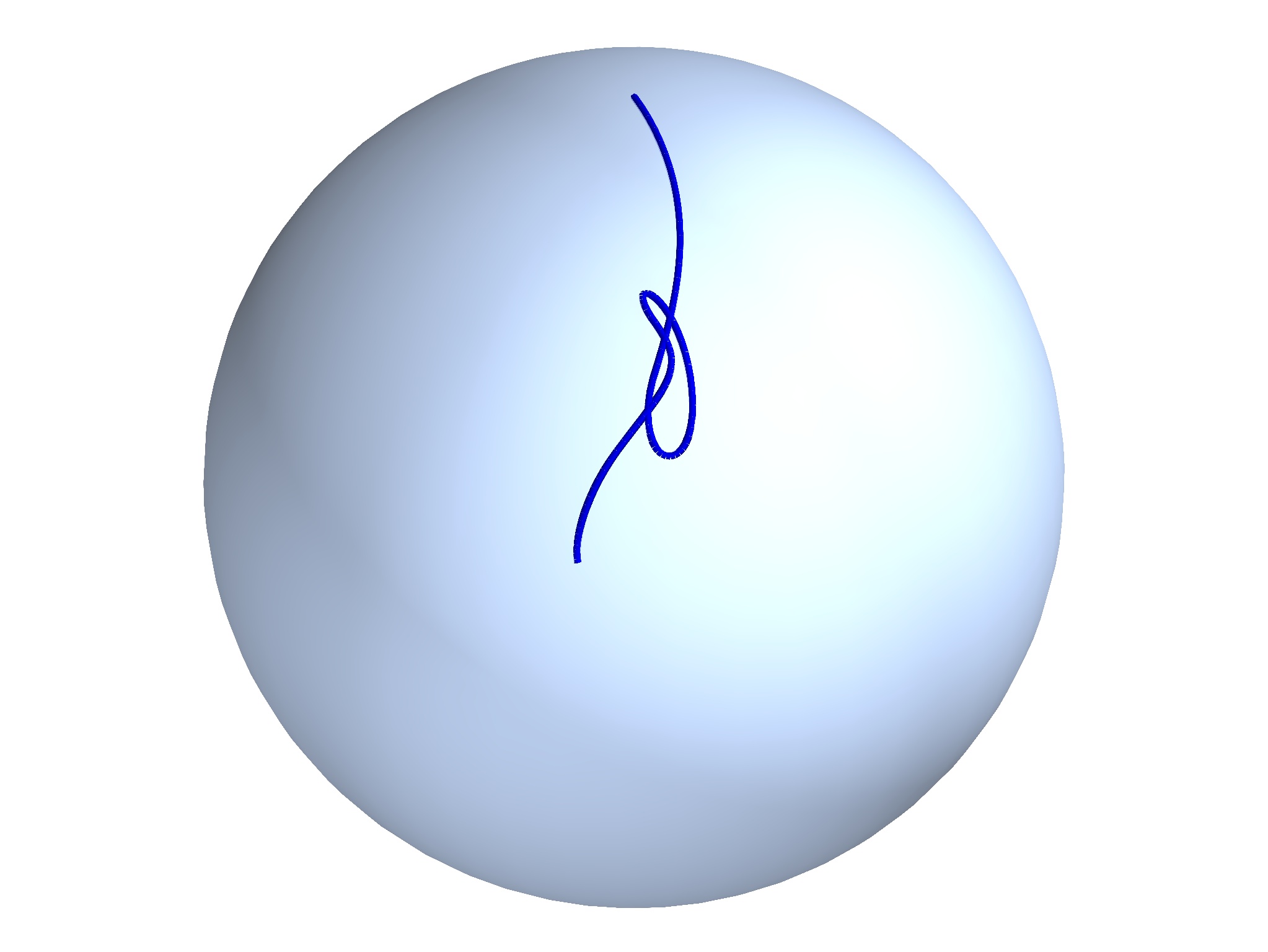}
\label{HCurves:subfig2}}

\subfigure[Field strength $\omega(t)$ for $\sigma = 0.04$]{
\includegraphics[scale=0.18]{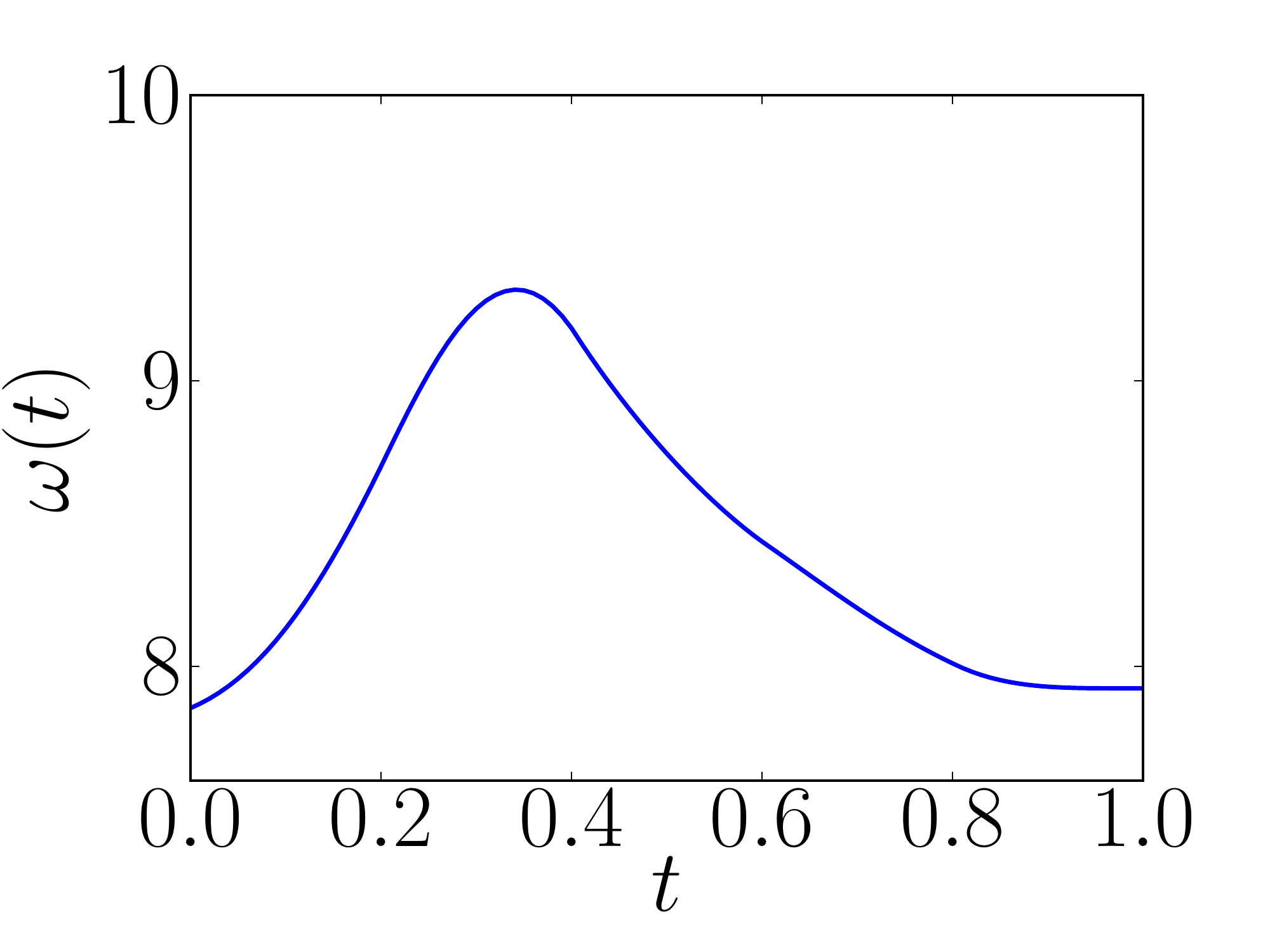}
\label{NCurves:subfig3}}
\qquad
\subfigure[Field strength $\omega(t)$ for $\sigma = 0.01$]{
\includegraphics[scale=0.18]{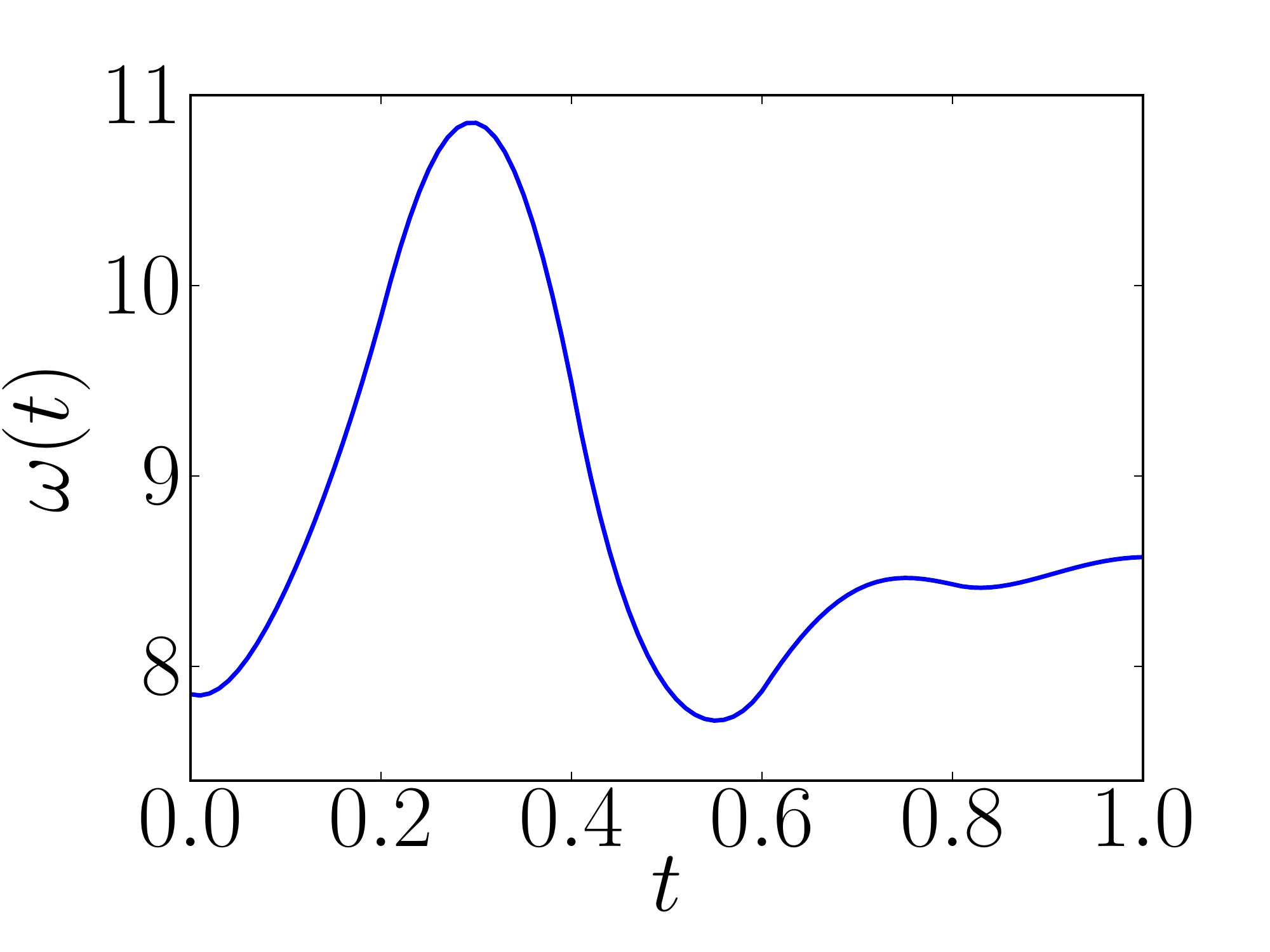}
\label{HCurves:subfig4}}
\caption{\footnotesize (Colour online) 
\textit{The quantum spline $H(t)$}. 
Hamiltonians that generate the dynamical trajectories in Fig.~\ref{fig:2}. 
The top row shows the orbits of the endpoint of the rotation axis ${\boldsymbol n}(t)$. 
The bottom row shows the field strength $\omega(t)$. These images illustrate the fact 
that as the value of $\sigma$ is decreased, the amount of change in the optimal Hamiltonian 
$H(t)$ increases. 
\label{fig:3}
}
\end{figure}

Before we indicate the process for the implementation of the optimisation scheme, let us 
show some results first. Consider a two-level system ($n=1$). We can think of this system 
as a spin-$\frac{1}{2}$ particle immersed in a magnetic field. If ${\boldsymbol n}(t)$ 
is the unit direction of the field at time $t$, the Hamiltonian of the system can be written in the 
form $H(t)=\omega(t) {\boldsymbol\sigma}\!\cdot\!{\boldsymbol n}(t)$, where $\omega(t)$ is 
the field strength. In this case the state space is just the Bloch sphere $S^2$. We have 
implemented the optimisation for a set of target states on $S^2$, an initial state $|\psi_0\rangle$, 
and a set of times. Using the resulting Hamiltonian we have 
generated the dynamics of the state, as illustrated in Fig.~\ref{fig:1}. In Fig.~\ref{fig:2} we 
have sketched the effect of choosing different tolerance levels. When the value of $\sigma$ is 
reduced, the resulting orbit $|\psi_t\rangle$ traverses closer to the vicinities of the 
target states $\{|\phi_j\rangle\}$. From (\ref{eq:3}), one sees that 
this may be realised at the expense of varying the Hamiltonian $H(t)$ more rapidly. 
This effect can be visualised in the case of a two-level system, since $H(t)$ is 
characterised by the the scalar field strength $\omega(t)$ and the unit vector 
${\boldsymbol n}(t)\in{\mathds R}^3$. In Fig.~\ref{fig:3} we have plotted the end-point 
of the unit vector 
${\boldsymbol n}(t)$ on a sphere, and the values of $\omega(t)$, for different choices of 
$\sigma$. These plots show that both ${\boldsymbol n}(t)$ and $\omega(t)$ vary 
more rapidly at smaller tolerance level (i.e. smaller $\sigma$). 

Another example we consider here is a controlled motion of a quantum state on the 
coherent-state subspace of the state space. Consider $SU(n+1)$ coherent states 
\cite{Per1972,Gil1972} in arbitrary dimensions. These coherent states can be generated 
by taking symmetric tensor products of `single-particle' states. In the context of quantum 
information theory, these states correspond to totally disentangled 
states inside the symmetric subspace of the Hilbert space of the combined system. Each 
coherent state thus corresponds to the image of a map, known as the \textit{Veronese 
embedding} \cite{BrGr2010,BrHu2001}, of a pure state. Therefore, given a set of points on a 
coherent-state space we identify them with states on a single-particle Hilbert space, solve 
the quantum spline problem as indicated above, and map the result back to the larger 
Hilbert space. In particular, the coherent quantum spline is generated by the symmetric 
tensor product Hamiltonian $\bigotimes_{\rm s}H(t)$. This elementary procedure 
works because (a) the Veronese embedding commutes with the action of 
$SU(n+1)$; and (b) the natural metrics on the spaces of coherent states are scalar-multiples 
of the metric (\ref{eq:2}) used here \cite{BrGr2010}. 

Next we discuss a numerical approach for finding a local minimum of the cost 
functional (\ref{eq:3}). The search can be restricted to solutions of (\ref{eq:5}) 
and (\ref{eq:9}), which are encoded by their initial conditions $M(t_0)$ and $\dot{H}(t_0)$. 
We can therefore regard \eqref{eq:3} as a function of these initial 
conditions, and perform a descent algorithm on that function. The terminal conditions at $t_m$ 
can then be used to test whether we have arrived at a local minimum. 

For a numerical implementation we can discretise the equations of motion (\ref{eq:5}) and 
(\ref{eq:9}), and find the approximate gradient of $\mathcal{J}$; alternatively, we can introduce 
an approximation $\mathcal{J}_d$ of $\mathcal{J}$ defined on a discrete path space, and take 
its variation, which yields a set of discrete equations of motion. Here we follow the latter method, 
which permits the use of adjoint equations~\cite{BHM} for an efficient calculation of the 
exact gradient of $\mathcal{J}_d$. This method is highly effective in dealing with 
higher-dimensional ($n > 1$) systems. Moreover, in this method discrete critical curves of 
${\mathcal J}_d$ satisfy a version of the terminal conditions at $t = t_m$ \emph{exactly}, and 
this leads to a precise method for testing convergence. In addition, such curves fulfil the 
conditions for the above-stated Lemma on their discrete time domain, which can be exploited 
by restricting the search for the optimal initial value of $M$ to $\mathcal{P}_{\psi_0}^\perp$. 

The implementation will make use of the \textit{Cayley map} $\tau:~\mathfrak{su}(n+1) \to 
SU(n+1)$, which approximates the Lie exponential according to $X \mapsto ({\mathds 1} - 
X/2)^{-1} ({\mathds 1} + X/2)$. We will also need the left-trivialised differential $d_l$:  
$d_l\tau_XY = (\rd/\rd\varepsilon) \tau(X + \varepsilon Y) \tau(X)^{-1}|_{\varepsilon = 0}$, 
which is given by $({\mathds 1} -X/2)^{-1}Y({\mathds 1} +X/2)^{-1}$.
We discretise the time interval $t_m-t_0$ into $N$ steps such that $(t_m-t_0)/N=h$, and 
we let $t^\mu = t_0 + \mu h$ for $\mu=0,\ldots,N$. For simplicity, we assume that the nodal 
times $\{t_j\}_{j=0,\ldots,m}$ coincide with some of the discrete time steps 
$t^{n_j}=t_0+n_j h$, where $n_0=0$ and $n_m=N$. To obtain a discrete version of 
the cost functional, we approximate the time derivative $-\ri\dot{H}$ of the generator 
by the discrete variables $\{L_\mu\}$. The complete set of discrete variables is 
therefore $(U_\mu, \ri H_\mu, M_\mu, L_\mu)$, with $\mu=0,\ldots,N$. Writing 
$\Delta_{\mu}=\delta_{\mu n_j}D_j F_j/\sigma^2$ and making 
use of the Euler method of \cite{BHM}, we obtain the following set of discrete equations 
of motion for $\mu=0,\ldots,N-1$:
\begin{eqnarray}
M_{\mu+1} &=& (d_l\tau^{-1}_{{\rm i} hH_{\mu+1}})(d_l\tau_{-{\rm i} hH_{\mu+1}}) 
(M_\mu + \Delta_\mu) \nonumber \\ 
L_{\mu+1} &=& L_\mu-h(d_l\tau_{{\rm i} hH_{\mu+1}})M_{\mu+1} \label{eq:11} \\ 
U_{\mu+1} &=& \tau(-{\rm i} hH_{\mu+1})U_\mu\,, \quad H_{\mu+1}=H_\mu+\ri hL_\mu\,. 
\nonumber 
\end{eqnarray}
These equations can be integrated for given initial values $M_0$ and $L_0$ (recall 
that $U_0={\mathds 1}$ and $H_0$ are prescribed). The terminal conditions are 
$L_N=0$ and $M_N +\Delta_N=0$. The discrete cost functional ${\mathcal J}_d$ 
in terms of initial 
conditions $(M_0, L_0)$ is
\begin{eqnarray}
{\mathcal J}_d = \sum_{\mu = 0}^{N-1} \frac{h}{2} \langle L_\mu, L_\mu \rangle 
+ \frac{1}{2\sigma^2} \sum_{j=1}^{m} D^2(U_{n_j}\psi_{0}, \phi_j), 
\label{eq:12}
\end{eqnarray}
whereby the equations of motion \eqref{eq:11} are implied.

A local minimum can be found by a gradient descent method, which requires the 
computation of the gradient of $\mathcal{J}_d$. The estimation of the gradient via 
finite-difference methods requires the repeated forward integration of the system of 
equations \eqref{eq:11}. The number of forward integrations increases with the 
number of dimensions of the Lie algebra ($n^2$ to leading order). Such estimation 
procedures thus quickly become unfeasible for higher-dimensional systems. This 
difficulty can be avoided by using the method of \emph{adjoint equations}, which can 
be readily implemented for the discretisation (\ref{eq:11}), (\ref{eq:12}) presented here 
(see Supplemental Material for details, and arXiv:1206.2675v2 for a numerical code). 
Then, the exact gradient is obtained at the cost of 
integrating \emph{twice} (once forward, once backward) a system of equations of the 
same complexity as (\ref{eq:11}). This allows for an efficient treatment of the quantum 
spline problem when $n>1$. 

We thank C. Burnett and L. Noakes for fruitful and thoughtful discussions in the course 
of this work. DDH and DMM are also grateful for partial support by a European Research 
Council Advanced Grant.


\begin{appendix}

\section{Gradient computation via adjoint equations: Supplemental Material}

Here we describe the method of \emph{adjoint equations} for the efficient computation of the 
gradient of  $\mathcal{J}_d$. We will supply the necessary equations, referring to \cite{BHM} 
for further details. First we need convenient expressions for the partial derivatives of 
$\mathcal{J}_d$ with respect to the variables $M_0$ and $L_0$. These are denoted by 
$\nabla_{M_0}\mathcal{J}_d$ and  $\nabla_{L_0}\mathcal{J}_d$, respectively, and are defined 
by
\begin{eqnarray}
 \delta \mathcal{J}_d &=& \left.\frac{\rd }{\rd \varepsilon} \mathcal{J}_d
 (M_0 + \varepsilon \delta M_0, L_0 + \varepsilon \delta L_0)\right|_{\varepsilon = 0} 
 \nonumber \\ &=& 
 \langle \nabla_{M_0}\mathcal{J}_d, \delta M_0\rangle + \langle \nabla_{L_0} 
 \mathcal{J}_d, \delta L_0\rangle\,,
\label{Gradient_def}
\end{eqnarray}
for all $\delta M_0, \,\delta L_0$ in $\mathfrak{su}(n+1)$.

Besides the Cayley map $\tau$ and the left-trivialised differential $d_l$ we will also need the 
right-trivialised differential $d_r$: $d_r\tau_XY = \tau(X)^{-1} (\rd/\rd\varepsilon) \tau(X + 
\varepsilon Y)|_{\varepsilon = 0}$, which is given by $({\mathds 1} +X/2)^{-1}Y({\mathds 1} 
-X/2)^{-1}$. We define the functions $F_j$ for $j = 1, \ldots m$: 
\begin{eqnarray}
F_j(\psi) = \frac{\langle\psi|\phi_j\rangle |\psi\rangle \langle\phi_j| -
\langle\phi_j|\psi\rangle |\phi_j\rangle \langle\psi|}
{\sin (D(\psi, \phi_j))\langle\phi_j|\phi_j\rangle \langle\psi|\psi\rangle},
\notag
\end{eqnarray}
as well as the functions $\Delta_\mu$ for $\mu = 0, \ldots, N$ defined by $\Delta_{\mu}(\psi) = 
\delta_{\mu n_j}D(\psi, \phi_j)F_j(\psi)/\sigma^2$.

Next, define an \emph{augmented functional} $\mathcal{G}$, in which the discrete 
equations of motion \eqref{eq:11} are incorporated using Lagrange multipliers. These Lagrange 
multipliers will be denoted $\{P^0_{\mu}, P^1_\mu, V^0_\mu, V^1_\mu\}_{\mu=1,\ldots,N}$. Let 
us introduce the shorthand notation $x$ representing the discrete path $\{U_\mu, \ri H_\mu, 
M_\mu, L_\mu\}_{\mu=0,\ldots,N}$ and $\lambda$ representing the Lagrange multipliers 
$\{P^0_{\mu}, P^1_\mu, V^0_\mu, V^1_\mu\}_{\mu=1,\ldots,N}$. Writing $|\psi_\mu\rangle = 
U_\mu|\psi_0\rangle$ the  augmented functional $\mathcal{G}$ is given by
\begin{eqnarray}
\mathcal{G}(x, \lambda) &=& h \sum_{\mu = 0}^{N-1} \bigg[\half \langle L_\mu, L_\mu\rangle 
+ \langle P^0_{\mu+1}, \tau^{-1}(U_{\mu+1}U_\mu^{-1}) \nonumber \\ && + \ri h H_{\mu+1}\rangle 
+ \langle P^1_{\mu+1}, \ri H_\mu -\ri H_{\mu+1} - hL_\mu\rangle \nonumber \\
& & + \langle V^0_{\mu+1}, L_{\mu+1} - L_\mu + h(d_l\tau_{{\rm i}hH_{\mu+1}})M_{\mu+1}\rangle 
\nonumber \\
& & + \langle V^1_{\mu+1},(d_l\tau_{{\rm i}hH_{\mu+1}}) M_{\mu+1} \nonumber \\ && 
-(d_l\tau_{-{\rm i} hH_{\mu+1}})\left(M_\mu + \Delta_\mu(\psi_\mu)\right)\rangle \bigg] 
\nonumber \\
& &  +  \frac{1}{2\sigma^2} \sum_{j=1}^{m} D^2(U_{n_j} \psi_0, \phi_j). \nonumber 
\end{eqnarray}
No constraints are assumed here, apart from the prescribed initial Hamiltonian $H_0$ and 
$U(0) = {\mathds 1}$. Note that $\mathcal{G}(x, \lambda) = \mathcal{J}_d(M_0, L_0)$ for any 
choice of Lagrange multipliers $\lambda$, provided that $x$ satisfies the 
discrete equations of motion \eqref{eq:11} for given initial values $M_0$ and $L_0$. 
Moreover, taking variations of $\mathcal{G}$ yields
\begin{align}
\label{DeltaG}
\delta \mathcal{G} = h \langle L_0 \!-\! h P^1_1 \!-\! V^0_1, \delta L_0\rangle  - 
h \langle d_r\tau_{-{\rm i}hH_1} \! V_1^1, \delta M_0 \rangle 
\end{align}
if $x$ satisfies \eqref{eq:11} \emph{and} $\lambda$ is a solution of the \emph{adjoint 
equations}. To specify the adjoint equations, we introduce functions $K^{\pm}$ by the 
defining relation
\begin{align}
\left.\frac{\rd}{\rd \varepsilon}\langle d_l
\tau_{\pm h(X + \varepsilon Y)} M, V \rangle \right|_{\varepsilon = 0}  = 
\langle K^{\pm}_{(X, M)}V, Y\rangle  \notag
\end{align}
for all $X, Y, M, V $ in $ \mathfrak{su}(n+1)$, and define $\mathcal{A}_\mu$ by
\begin{align}
\left.\frac{\rd}{\rd \varepsilon}\langle\Delta_\mu\left(\re^{\varepsilon A}\psi\right), V \rangle 
\right|_{\varepsilon = 0}  = \langle \mathcal{A}_\mu(\psi, V), A\rangle \notag
\end{align}
for all  $V, A \in \mathfrak{su}(n+1)$ and state vectors $|\psi\rangle$. The \emph{adjoint 
equations} consist of conditions at the final time point: 
\begin{align}
&V_N^0 = 0, \quad V_N^1 = 0, 
\label{Adj_N_1} \\
&P_N^0 = -\frac{1}{h} (d_l\tau_{-ihH_N})\Delta_N(\psi_N), \quad P_N^1 = hP^0_N, 
\label{Adj_N_2}
\end{align}
and the following set of equations
\begin{align}
  &V_\mu^0 = V^0_{\mu+1} + hP^1_{\mu+1} - L_\mu \notag\\
  &V_\mu^1 = d_r\tau^{-1}_{{\rm i}hH_\mu}\left[ d_r\tau_{-{\rm i}hH_{\mu+1}}V^1_{\mu+1} - 
  h d_r\tau_{{\rm i}hH_\mu}V_\mu^0\right] \notag\\
  &P^0_\mu = d_l\tau_{{\rm i}hH_\mu} \Big[d_l\tau^{-1}_{-{\rm i}hH_{\mu+1}} P^0_{\mu+1} 
  - \frac{1}{h} \Delta_\mu(\psi_\mu)
  \nonumber \\ & 
  \qquad  \qquad \qquad 
  + \mathcal{A}_\mu(\psi_\mu, d_r\tau_{-{\rm i}hH_{\mu+1}} V^1_{\mu+1}) 
  \Big] \label{Ad_3}\\
&P^1_\mu = P^1_{\mu+1} + h P^0_\mu - hK^-_{(-{\rm i}H_\mu, M_\mu)}
V^0_\mu - K^-_{(-{\rm i}H_\mu, M_\mu)}V_\mu^1 \notag \\ 
& \qquad + K^+_{(-{\rm i}H_\mu, M_{\mu-1} + \Delta(\psi_{\mu-1}))}V^1_\mu \notag\,,
\end{align}
for $\mu = 1, \ldots, N-1$. These equations are posed backwards. That is, solving the adjoint 
equations entails initialising the Lagrange multipliers at time point $N$ according to 
\eqref{Adj_N_1} and \eqref{Adj_N_2}, and then iterating backwards from $\mu = N$ to 
$\mu = 1$ using  \eqref{Ad_3}.

We now obtain the exact gradient of $\mathcal{J}_d$ from \eqref{DeltaG}. Indeed, let 
$(M_0(\varepsilon), L_0(\varepsilon))$ be variations of initial conditions $(M_0,L_0)$, 
let $x(\varepsilon)$ be the corresponding solutions to the discrete equations of motion 
\eqref{eq:11} with $x = x(0)$, and let $\lambda$ be a solution to the adjoint equations 
\eqref{Adj_N_1}--\eqref{Ad_3}. Then
\begin{align}
  \delta \mathcal{J}_d &= \left.\frac{\rd}{\rd \varepsilon} \mathcal{J}_d(M_0(\varepsilon), 
  L_0(\varepsilon))\right|_{\varepsilon = 0} =  \left.\frac{\rd}{\rd \varepsilon} \mathcal{G}
  (x(\varepsilon), \lambda)\right|_{\varepsilon = 0} \notag \\
&=  h \langle L_0 - h P^1_1 - V^0_1, \delta L_0\rangle  - h \langle d_r\tau_{-{\rm i}hH_1}
V_1^1, \delta M_0 \rangle, \notag
\end{align}
where we have used \eqref{DeltaG} in the last equality. Recalling definition \eqref{Gradient_def} 
we conclude that 
\begin{align}\label{Gradient_expression}
\nabla_{M_0}\!\mathcal{J}_d =  - h D\tau_{-{\rm i}hH_1}V_1^1, \quad \!\!\!\! 
\nabla_{L_0}\!\mathcal{J}_d =   h\left( L_0 - h P^1_1 - V^0_1\right).
\end{align}

In summary, the partial derivatives $\nabla_{M_0}\mathcal{J}_d$ and 
$\nabla_{L_0}\mathcal{J}_d$ are computed as follows: (i) Integrate the system of equations 
\eqref{eq:11} up to $\mu = N$; (ii) Initialise the Lagrange multipliers at $\mu=N$ according to 
\eqref{Adj_N_1} and \eqref{Adj_N_2}; (iii) Integrate backwards the system of equations 
\eqref{Ad_3} until $\mu=1$; and (iv) Obtain $\nabla_{M_0}\mathcal{J}_d$ and 
$\nabla_{L_0}\mathcal{J}_d$ by evaluating the right hand sides of \eqref{Gradient_expression}.

\end{appendix}


\vspace{-0.5cm} 

\begin{references}


\bibitem{QC}
At the time of drafting this paper there are 935 papers posted on the arXiv 
that contain the terms ``quantum'' and ``control'' in the title. 

\bibitem{DCB} Brody,~D.~C. 
``Elementary derivation for passage times'' 
{\em J. Phys.} A\textbf{36}, 5587--5593 (2003). 

\bibitem{Hosoya} Carlini,~A., Hosoya,~A., Koike,~T. \& Okudaira,~Y. 
``Time-optimal quantum evolution'' 
{\em Phys. Rev. Lett.} \textbf{96}, 060503 (2006). 

\bibitem{BH} Brody,~D.~C. \& Hook,~D.~W. 
``On optimum Hamiltonians for state transformations'' 
{\em J. Phys.} A\textbf{39}, L167--L170 (2006); Corrigendum 
\textit{ibid}. \textbf{40}, 10949 (2007).

\bibitem{GHM} Gay-Balmaz,~F., Holm,~D.~D., Meier,~D.~M., 
Ratiu,~T.~S., \& Vialard,~F.~X.  
``Invariant higher-order variational problems'' 
{\em Commun. Math. Phys.} {\bf 309}, 423--458 (2012).

\bibitem{BHM} Burnett, ~C., Holm,~D.~D.,  Meier,~D.~M.
``Geometric integrators for higher-order mechanics on Lie groups'' (2012)
Preprint available at \url{http://arxiv.org/abs/1112.6037}

\bibitem{NHP1989}
Noakes,~L., Heinzinger,~G., \& Paden,~B.
\newblock {``Cubic splines on curved spaces''}
\newblock {\em IMA Journal of Mathematical Control \& Information} {\bf 6}, 
465--473 (1989).

\bibitem{Per1972}
Perelomov,~A.~M.
``Coherent states for arbitrary Lie groups''
{\em Commun. Math. Phys.} \textbf{26}, 222--236 (1972).

\bibitem{Gil1972}
Gilmore,~R.
``Geometry of symmetrised states''
{\em Ann. Phys.} \textbf{74}, 391--463 (1972).

\bibitem{BrGr2010} Brody,~D.~C. \& Graefe,~E.~M.
``Coherent states and rational surfaces''
{\em J. Phys.} A\textbf{43}, 255205--255219 (2010). 

\bibitem{BrHu2001} Brody,~D.~C. \& Hughston,~L.~P.
``Geometric quantum mechanics''
{\em J. Geom. Phys.} \textbf{38}, 19--53 (2001).

\end{references}
\end{document}